\documentclass{article}

\usepackage{amsthm}
\usepackage{graphicx}

\begin{document}  

\title{TESTING NEW PROPERTY OF ELLIPTICAL MODEL FOR STOCK RETURNS DISTRIBUTION.}

\author{Petr Koldanov}






\maketitle


\begin{abstract}
Wide class of elliptically contoured distributions is a popular model of stock returns distribution. However the important question of adequacy of the model is open. There are some results which reject and approve such model. Such results are obtained by testing some properties of elliptical model for each pair of stocks from some markets. New property of equality of $\tau$ Kendall correlation coefficient and probability of sign coincidence for any pair of random variables with elliptically contoured distribution is proved in the paper. Distribution free statistical tests for testing this property for any pair of stocks are constructed. Holm multiple hypotheses testing procedure based on the individual tests is constructed and applied for stock markets data for the concrete year. New procedure of testing the elliptical model for stock returns distribution for all years of observation for some period is proposed.
 The procedure is applied for the stock markets data of China, USA, Great Britain and Germany for the period from 2003 to 2014.
It is shown that for USA, Great Britain and Germany stock markets the hypothesis of elliptical model of stock returns distribution could be accepted but for Chinese stock market is rejected for some cases.  
 

\end{abstract}


\section{Introduction}	

Models of multivariate stock returns distribution attract growing attention last decade. In particular such model are needed for portfolio construction and risk management. Popular model of joint stock return distribution is wide class of elliptically contoured distributions (see \cite{Anderson_2003},\cite{Gupta_2013}). 
Consistency of the elliptical model with US and Japanese stock market data was studied in \cite{Chicheportiche_2012} where it was shown that joint distribution of stock return is not elliptical. Such result was obtained by comparison of  pairwise dependence measures between any pair of stocks. 
At the same time it was noted by authors that applied methodology is not statistical. 

In \cite{Koldanov_2016} statistical methodology to testing symmetry of stock returns distribution was proposed. 
Distribution free tests for testing symmetry of any pair of stock returns distribution was constructed. 
These individual tests using well known Holm procedure \cite{Holm_1979} was combined to testing symmetry of joint distribution of stock returns for US and UK markets. The concept of rejection graph was introduced and it was shown that deleting hubs (vertices with high degree) of the graph lead to acceptance of symmetry hypothesis of stock return distribution.  
    
The present work continues the studies begun in \cite{Koldanov_2016}. 
New property of elliptically contoured distributions namely the property of equality of $\tau$ Kendall correlation coefficient and probability of sign coincidence (measure ${\cal Q}$(\cite{Kruskal1958})) for any pair of random variables is proved in the paper. This property is used for testing elliptical model of stock returns distribution. 
Distribution free tests for individual hypotheses testing $h_{i,j}:\tau_{i,j}={\cal Q}_{i,j}$ against $k_{i,j}:\tau_{i,j}\neq {\cal Q}_{i,j}$ for any $i,j=1,\ldots,p$ (where $p$ is the number of stocks in the market) are constructed.
These individual tests using Holm procedure are combined to test consistency of elliptical model with joint stock return distribution. 
The concept of rejection graph \cite{Koldanov_2016} is used to describe results of the Holm procedure. Obtained results shows that rejection of elliptical model is connected with small number of pairs of stocks from Chinese stock market. Removing this stocks leads to nonrejection of elliptical model for stock return distribution.   

New procedure of testing the elliptical model for stock returns distribution for some period of years is constructed. The procedure is based on combination of the results of the Holm procedures for different years. The new procedure is applied for the stock markets data of China, USA, Great Britain and Germany for the period from 2003 to 2014. It is shown that for USA, Great Britain and Germany stock markets the hypothesis of elliptical model of stock returns distribution could be accepted. In contrast it is shown that for Chinese stock market the hypothesis of elliptical model of stock returns distribution is rejected for some cases.

The paper is organized as follows: in section \ref{Basic_definitions_and_notations} main notations and definitions are introduced and theorems of equality of $\tau$-Kendall correlation coefficient and measure ${\cal Q}$ (\cite{Kruskal1958}) for elliptically contoured distributions are proved. 
In section \ref{Tests_of_individual_hypotheses} individual tests are constructed.
In section \ref{Holm_procedure} Holm procedure is described.
In section \ref{experimental_result} the new procedure is presented and obtained results of experiments are described.
In section \ref{Concluding_remarks} discussion of obtained results is presented.
 
\section{Pairwise measures for elliptically contoured distributions}\label{Basic_definitions_and_notations}
Let $\left(\begin{array}{l}X_1\\X_2\\ \ldots\\X_p\end{array}\right)$ be the continuous random vector.
There are several correlation coefficient (or pairwise measures of dependence) between random variables $X_i,X_j$ such as Pearson correlation, Spearman correlation, Kendall correlation, Fechner correlation, Kruscal correlation. Some of them was investigated in \cite{Kruskal1958}. Kruscal correlation is based on the following measure.

{\bf Definition 1}\label{Kruscal_measure}

Measure ${\cal Q}$ of dependence between random variables $X_i$ and $X_j$ (\cite{Kruskal1958}) is
${\cal Q}_{i,j}=P\left[(X_i-med(X_i))(X_j-med(X_j))>0\right]$
where $med(X_i)$ is the median of the distribution $F_{X_i}(x)$ of the random variable $X_i$ i.e. $P(X_i>med(X_i))=P(X_i<med(X_i))=\frac{1}{2}$ or $F_{X_i}(med(X_i))=\frac{1}{2}$. 

Also it was shown that $\tau$-Kendall correlation coefficient $\hat{\tau}_{i,j}$ is based on the following measure:

{\bf Definition 2}\label{tau_Kendalla}

Let $\left(\begin{array}{l}X_i\\ X_j\end{array}\right)$ be random vector with cumulative distribution function $F_{X_iX_j}(x,y)$ and let $\left(\begin{array}{l}X_i(t)\\ X_j(t)\end{array}\right)$, $\left(\begin{array}{l}X_i(t+1)\\ X_j(t+1)\end{array}\right)$ be the independent copies of the vector $\left(\begin{array}{l}X_i\\ X_j\end{array}\right)$.
  
Kendall measure $\tau_{i,j}$ of dependence between random variables $X_i$ and $X_j$ is defined by equality 
$$\tau_{i,j}=P\left[(X_i(t)-X_i(t-1))(X_j(t)-X_j(t-1))>0\right]$$

It was proved that if random vector $X=(X_1,X_2,\ldots,X_p)$ has multivariate normal distribution $N(\mu,\Lambda)$ with known $\mu$
then one has 
\begin{equation}\label{equality_kendall_tau_normal}
{\cal Q}_{i,j}=\tau_{i,j}\mbox{ for all }i,j=1,\ldots,p;i\neq j
\end{equation}

Let us prove that (\ref{equality_kendall_tau_normal}) is true for elliptically contoured distributions $ECD(\mu,\Lambda,g)$ with any function $g$.

Random vector $X=(X_1,X_2,\ldots,X_p)$ has elliptically contoured distribution $ECD(\mu,\Lambda,g)$ if it's density has the form: 
\begin{equation}\label{density _of_elliptical_distribution} 
f(x; \mu, \Lambda)=|\Lambda|^{-\frac{1}{2}}g\{(x-\mu)'\Lambda^{-1}(x-\mu)\}
\end{equation}
where $\Lambda$ is a positive definite matrix, function $g(x)\geq 0$ and 
$$
\int_{-\infty}^{\infty}\ldots\int_{-\infty}^{\infty}g(y'y)dy_1\ldots dy_p=1
$$
The class of elliptically contoured distributions contain in particular multivariate normal distribution and multivariate Student distribution. It is known that $E(X_i)=\mu_i$, $i=1,2,\ldots, p$ if it exist. 

{\bf Theorem 1}

If random vector $X$ has elliptically contoured distribution $ECD(\mu,\Lambda,g)$ with known $\mu$ then for any $g$ one has ${\cal Q}_{i,j}=\tau_{i,j}, \forall i,j=1,\ldots,p, i\neq j $ .

{\bf Proof}

Vector $X$ has elliptically contoured distribution $ECD(\mu,\Lambda,g)$ then $med(X_i)=\mu_i$, i.e. vector $U=X-med(X)$ has elliptically contoured distribution $ECD(0,\Lambda,g)$. 
Vectors $X(t),X(t-1)$ are independent (by definition 2) and have elliptically contoured distribution $ECD(\mu,\Lambda,g)$ then by lemma 1 from \cite{Lindskog_2003} vector $V=X(t)-X(t-1)$ has elliptically contoured distribution $ECD(0,\Lambda,g_1)$ where 
$$
\int_{-\infty}^{\infty}\ldots\int_{-\infty}^{\infty}g_1(y'y)dy_1\ldots dy_p=1
$$

In \cite{KKK2017}, lemma 2 it was shown that if $X$ has elliptically contoured distribution $ECD(0,\Lambda, g)$ then $P(X_i>0,X_j>0)=\frac{1}{4}+\frac{1}{2\pi}\arcsin\left(\frac{\lambda_{ij}}{\sqrt{\lambda_{ii}}\lambda_{jj}}\right)$ (where $\lambda_{ij}$ are the elements of the matrix $\Lambda$) and does not depend from $g$. 

Therefore $P(U_iU_j>0)=P(V_iV_j>0)$ or ${\cal Q}_{i,j}=\tau_{i,j}$ for $\forall g$. The theorem 1 is proved.

The theorem 1 has the following

{\bf Corollary 1}

If vector $\mu$ is known and exists $i,j=1,\ldots,p;i\neq j$ such that ${\cal Q}_{i,j}\neq\tau_{i,j}$ then the vector $X$ has not  elliptically contoured distribution.  

In real practice it is unrealistic to assume known vector $\mu$. To deal with unknown $\mu$ let us prove the following

{\bf Theorem 2}

Let 
$\left(
				\begin{array}{l}
				  X_1(1)\\
					X_2(1)\\
					\ldots\\
					X_p(1)\\
				\end{array}
		\right)
,\left(
				\begin{array}{l}
				  X_1(2)\\
					X_2(2)\\
					\ldots\\
					X_p(2)\\
				\end{array}
		\right)
,\ldots,
\left(
				\begin{array}{l}
				  X_1(n)\\
					X_2(n)\\
					\ldots\\
					X_p(n)\\
				\end{array}
		\right)
$  be the sample of independent identically distributed observations from vector 
$X=\left(
				\begin{array}{l}
				  X_1\\
					X_2\\
					\ldots\\
					X_p\\
				\end{array}
		\right)
$ which has elliptically contoured distribution $ECD(\mu,\Lambda,g)$. Let 
$$\overline{X_i}=\frac{1}{n}\sum_{t=1}^nX_i(t)$$
Then for $n\geq 2$ one has 
$$P((X_i(t)-\overline{X_i})>0,(X_j(t)-\overline{X_j})>0)=\frac{1}{4}+\frac{1}{2\pi}\arcsin \frac{\lambda_{ij}}{\sqrt{\lambda_{ii}\lambda_{jj}}},\forall i,j=1,\ldots,p;\forall t=1,\ldots,n
$$ 		
{\bf Proof}

According to lemma 1 from \cite{Lindskog_2003} if vector $Z_1$ has elliptically contoured distribution $ECD(\mu_1,\Lambda,g_1)$ and vector $Z_2$ has elliptically contoured distribution $ECD(\mu_2,\Lambda,g_2)$, then vector $aZ_1+bZ_2$ has elliptically contoured distribution $ECD(a\mu_1+b\mu_2,\Lambda,g_3)$. Without loss of generality, consider the case $t=1$. One has 
$$X(1)-\overline{X}=\frac{n-1}{n}X(1)-\frac{1}{n}\sum_{t=2}^nX(t)$$
From lemma 1 of \cite{Lindskog_2003} vector 
$$\frac{1}{n}\sum_{t=2}^nX(t) \mbox{ has elliptically contoured distribution }ECD(\frac{n-1}{n}\mu,\Lambda,g^{\star})$$
Since vectors 
$X(1)$ and $\frac{1}{n}\sum_{t=2}^nX(t)$ are independent then vector   
$X(1)-\overline{X}$ has elliptically contoured distribution $ECD(0,\Lambda,g^{\star\star})$

From lemma 2 of \cite{KKK2017} one has
$$P((X_i(t)-\overline{X_i})>0,(X_j(t)-\overline{X_j})>0)=\frac{1}{4}+\frac{1}{2\pi}\arcsin \frac{\lambda_{ij}}{\sqrt{\lambda_{ii}\lambda_{jj}}},\forall t=1,\ldots,n;\forall i,j=1,\ldots, p,i\neq j$$
The theorem 2 is proved.

It follows from theorem 2 that if vector X has elliptically contoured distribution $ECD(\mu,\Lambda,g)$ then  
$$P((X_i(t)-\overline{X_i})>0,(X_j(t)-\overline{X_j})>0)=P((X_i(t)-\mu_i)>0,(X_j(t)-\mu_j)>0)={\cal Q}_{i,j}$$

From theorem 2 and corollary 1 one has the following

{\bf Corollary 2}

If exist $i,j=1,\ldots,p;i\neq j$ such that ${\cal Q}_{i,j}\neq\tau_{i,j}$ then the vector $X$ has not  elliptically contoured distribution.  
  
\section{Individual hypotheses testing}\label{Tests_of_individual_hypotheses}

As follows from corollary 2 for study consistency with elliptical model it is necessary to test the equality ${\cal Q}_{i,j}=\tau_{i,j}$ for any $i,j=1,\ldots,p;i\neq j$. 
This problem can be formulated as problem of simultaneous testing of the following hypotheses 
\begin{equation}\label{individual_hypotheses}
h_{i,j}:{\cal Q}_{i,j}=\tau_{i,j}\mbox{ against }k_{i,j}:{\cal Q}_{i,j}\neq\tau_{i,j}
\end{equation}

In the present section distribution free tests are constructed for testing individual hypotheses (\ref{individual_hypotheses}).  

Without loss of generality let us consider the case $i=1,j=2$.

Let  
$$\left(\begin{array}{l}x_1(1)\\x_2(1)\end{array}\right),\left(\begin{array}{l}x_1(2)\\x_2(2)\end{array}\right),\ldots,\left(\begin{array}{l}x_1(n+m)\\x_2(n+m)\end{array}\right)$$ be the sample from random vector 
$\left(\begin{array}{l}X_1\\X_2\end{array}\right)$.

For testing hypothesis $h_{1,2}:{\cal Q}_{1,2}=\tau_{1,2}$ against $k_{1,2}:{\cal Q}_{1,2}\neq\tau_{1,2}$ let us consider the statistics
\begin{equation}\label{statistic_tau_kendalla}
\hat{\tau}_{1,2}=\sum_{i=1}^{\left[\frac{n}{2}\right]}S\{x_1(2i)-x_1(2i-1))(x_2(2i)-x_2(2i-1)\}
\end{equation}

and 
\begin{equation}\label{statistics_kruscal_measure}
\hat{{\cal Q}}_{1,2}=\sum_{i=n+1}^{n+m}S((x_1(i)-\overline{x_1})\times(x_2(i)-\overline{x_2}))
\end{equation}
where
$$S(x)=\left\{\
							\begin{array}{ll}
							 1,&x>0\\
							 0,&x\leq 0
							\end{array}
				\right.			
$$
$$
\overline{x_k}=\frac{1}{m}\sum_{i=n+1}^{n+m}x_k(i), k=1,2
$$

Statistic $\hat{\tau}_{1,2}$ has binomial distribution $b(r,\tau_{1,2})$ where $r=\left[\frac{n}{2}\right]$, statistic $\hat{{\cal Q}}_{1,2}$ has binomial distribution $b(m,{\cal Q}_{1,2})$.
Moreover statistics $\hat{\tau}_{1,2}$, $\hat{{\cal Q}}_{1,2}$ are independent since they are based on different observations. Therefore 
$$
\begin{array}{l}
P(\hat{\tau}_{1,2}=k,\hat{{\cal Q}}_{1,2}=l)=C_r^k\tau_{1,2}^k(1-\tau_{1,2})^{r-k}C_m^l{\cal Q}_{1,2}^l(1-{\cal Q}_{1,2})^{m-l}=\\
=C_r^k C_m^l\exp\left\{\ k\ln(\frac{\tau_{1,2}}{1-\tau_{1,2}})+l\ln(\frac{{\cal Q}_{1,2}}{1-{\cal Q}_{1,2}})\right\}\ (1-\tau_{1,2})^r(1-{\cal Q}_{1,2})^m\\
\end{array}
$$
 
Then uniformly most powerful unbiased (UMPU) test ( see \cite{Lehmann2005}, p. 126) for testing $h_{1,2}:{\cal Q}_{1,2}=\tau_{1,2}$ against $k_{1,2}:{\cal Q}_{1,2}\neq\tau_{1,2}$ has the form:
\begin{equation}\label{UMPU_test_tau_kruscal}
\varphi_{1,2}(x)=\left\{\
									\begin{array}{ll}
									 0,& c_1(l+k)\leq k\leq c_2(l+k)\\
									 1,& k<c_1(l+k)\mbox{ or } k>c_2(l+k)
									\end{array}
						\right.
\end{equation}
where $c_1(l+k), c_2(l+k)$ are defined from:
\begin{equation}\label{threshold_equation}
 P_{{\cal Q}_{1,2}=\tau_{1,2}}(k<c_1(l+k)\mbox{ or } k>c_2(l+k)/\hat{\tau}_{1,2}+\hat{{\cal Q}}_{1,2}=l+k)=\alpha
\end{equation}
For simplicity let us consider the test $\varphi_{1,2}(x)$ where constants $c_1(l+k), c_2(l+k)$ are defined from equations:
\begin{equation}\label{equitail_equations_fir_constant}
\begin{array}{l}
P_{{\cal Q}_{1,2}=\tau_{1,2}}(k<c_1(l+k)/\hat{\tau}_{1,2}+\hat{{\cal Q}}_{1,2}=l+k)=\frac{\alpha}{2}\\
P_{{\cal Q}_{1,2}=\tau_{1,2}}(k>c_2(l+k)/\hat{\tau}_{1,2}+\hat{{\cal Q}}_{1,2}=l+k)=\frac{\alpha}{2}\\
\end{array}
\end{equation}
Since
$$
\begin{array}{l}
P_{{\cal Q}_{1,2}=\tau_{1,2}}(\hat{\tau}_{1,2}=k/\hat{\tau}_{1,2}+\hat{{\cal Q}}_{1,2}=l+k)=\frac{C_r^kC_m^l}{C_{r+m}^{k+l}}
\end{array}
$$
then $c_1(l+k)$ is greatest integer number satisfying: 
\begin{equation}\label{threshold_for_individual_test1}
\sum_{i=0}^{c_1(l+k)}\frac{C_r^i C_m^{l+k-i}}{C_{r+m}^{k+l}}\leq\frac{\alpha}{2}
\end{equation} 
$c_2(l+k)$ is smallest integer number satisfying: 
\begin{equation}\label{threshold_for_individual_test}
\sum_{i=c_2(l+k)}^{r}\frac{C_r^i C_m^{l+k-i}}{C_{r+m}^{k+l}}\leq\frac{\alpha}{2}
\end{equation} 
The p-value of the test (\ref{UMPU_test_tau_kruscal}) is defined as
\begin{equation}\label{p_value}
q_{1,2}=2\min\left\{\ \sum_{i=0}^k\frac{C_r^i C_m^{l+k-i}}{C_{r+m}^{k+l}}, \sum_{i=k}^r\frac{C_r^i C_m^{l+k-i}}{C_{r+m}^{k+l}}\right\}\
\end{equation}

\section{Holm procedure}\label{Holm_procedure}
In the case of large stock market the number of pairs of stocks is huge and it is necessary to take into account so called multiplicity phenomenon (\cite{Bretz_2011}). To study consistency of joint stock returns distribution with elliptical model the multiple hypotheses testing approach is used (see \cite{Lehmann2005}, ch.9).  Multiple comparison procedures adjust statistical inferences from an experiment for multiplicity and thus enable better decision making.  In our approach significance level of multiple test is given by the probability of at least one Type I error which is known  as Family Wise Error Rate (FWER).

The tests (\ref{UMPU_test_tau_kruscal}) with p-values from (\ref{p_value}) for testing the individual hypotheses (\ref{individual_hypotheses}) are combined in simultaneous testing procedure. In order to control FWER which is probability of at least one false rejection of true individual hypothesis (\ref{individual_hypotheses}) the well known Holm procedure \cite{Holm_1979} is used.

In our case Holm procedure contain at most $M=C_p^2$ steps. At any step one individual hypothesis $h_{i,j}$ is rejected or all other hypotheses are accepted. 
Holm procedure is defined by the following algorithm:

\begin{itemize}
\item Step 1: If  
$$\min_{i,j=1,\ldots,p}q_{i,j}\geq \frac{\alpha}{M} 
$$ 
then all hypotheses
$h_{i, j}, i, j=1,2,\ldots,p$ are accepted,  
 
 else if  $\min_{i, j=1,\ldots,p}q_{i, j}=  q_{i_1, j_1}$ 
 then hypothesis $h_{i_1, j_1}$ is rejected and go to the step 2.



\item \ldots
  
\item Step K: Let $I=\{(i_1, j_1), (i_2, j_2),\ldots,(i_{K-1}, j_{K-1})\}$ be the set of indexes of previously rejected hypotheses. 
If  $$\min_{(i, j)\notin I}q_{i, j}\geq \frac{\alpha}{M-K+1} $$ 
then accept the hypotheses
$h_{i, j}$, $(i, j)\notin I$, 

else if $\min_{(i, j)\notin I}q_{i, j}=q_{i_K,j_K}$ then reject hypothesis $h_{i_K, j_K}$ 
 and go to the step (K+1). 

\item \ldots

\item Step M: Let $I=\{(i_1, j_1),\ldots,(i_{M-1}, j_{M-1})\}$ be the set of indexes of previously rejected hypotheses. Let  $(i_M, j_M) \notin I$. 
If 
$$
q_{i_M, j_M}\geq \alpha $$ 
then accept hypothesis $h_{i_M, j_M}$, 
else reject hypothesis $h_{i_M, j_M}$ (reject all hypotheses).
\end{itemize}


\section{Testing of elliptical model. Experimental results}\label{experimental_result}

Our experimental results are presented in the section. First we present results for Chinese stock market which shows that hypotheses of elliptical model for stock returns could be rejected for the years 2006, 2010, 2011 for small number of stocks only.
Then we consider stock markets of different countries for some period and propose the procedure to test the hypothesis of elliptical model for stock returns distribution for the whole period. In our experimental results we consider stock markets of China, USA, Great Britain and Germany for the period from 2003 to 2014 years. 

\subsection{Testing of elliptical model for Chinese stock market. One year period}

To obtain experimental results for Chinese stock market 100 most traded stocks from Chinese stock market for 2006 year was selected. The Holm procedure with FWER=0.05; 0.5 and individual tests (\ref{UMPU_test_tau_kruscal}) with p-values from (\ref{p_value}) was applied to selected stocks for every  year from the period from 2003 to 2014. 

To describe the results of Holm procedure the concept of {\it rejection graph} introduced in \cite{Koldanov_2016} was used. Edge $(i,j)$ was added in the rejection graph if and only if the individual hypothesis $h_{i,j}:{\cal Q}_{i,j}=\tau_{i,j}$ was rejected by Holm multiple testing procedure, nodes of the rejection graph were vertices adjacent to this edge. 

In all presented results the values $n=m=125$ was selected. By the observations from the first half of a year the $\tau$ Kendall measure (Definition 2) was estimated by the observations from the second half of a year the $\cal{Q}$ Kruscall measure (Definition 1) was estimated. 

\begin{figure}
\includegraphics[scale=.5]{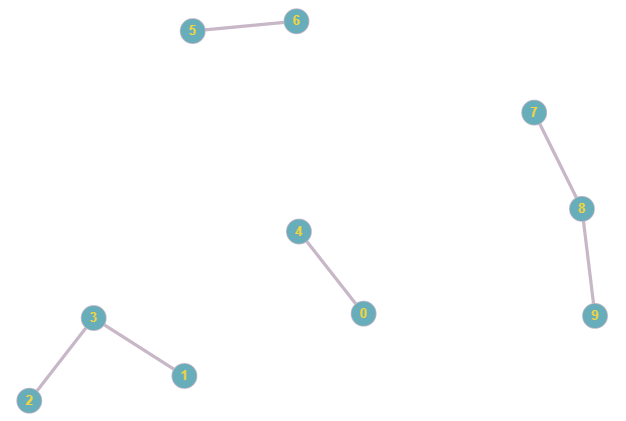}
\caption{Rejection graph for Chinese market 2006 year. FWER=0.5. $n=m=125$ }
\label{Chinese_market_2006}
\end{figure}

\begin{figure}
\includegraphics[scale=.6]{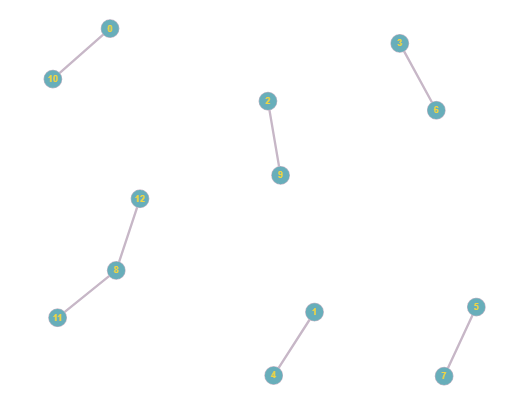}
\caption{Rejection graph for Chinese market 2010 year. FWER=0.5. $n=m=125$}
\label{Chinese_market_2010}
\end{figure}

The obtained results are shown at the figures \ref{Chinese_market_2006} (year 2006) and \ref{Chinese_market_2010} (year 2010). 
In the figures it is shown only verteces which are incident to some edges in the rejection graph. 
Note that in contrast to the rejection graph constructed in \cite{Koldanov_2016} in this case there are no hubs. At the same time the number of rejected hypotheses is very small and much less than in \cite{Koldanov_2016}.

The list of stock tickers from these figures is as follows:

\begin{itemize} 
\item Figure \ref{Chinese_market_2006}: vertex 0 correspond the ticker '600037.SS', vertex 1 correspond the ticker '600016.SS', vertex 2 correspond the ticker '600031.SS', vertex 3 correspond the ticker '600015.SS', vertex 4 correspond the ticker '600028.SS', vertex 5 correspond the ticker '600435.SS', vertex 6 correspond the ticker '600519.SS', vertex 7 correspond the ticker '600060.SS', vertex 8 correspond the ticker '600252.SS', vertex 9 correspond the ticker '600649.SS'
\item Figure \ref{Chinese_market_2010}: vertex 0 correspond the ticker '601555.SS',
   vertex 1 correspond the ticker '600029.SS',
   vertex 2 correspond the ticker '600037.SS',
   vertex 3 correspond the ticker '600031.SS',
   vertex 4 correspond the ticker '600795.SS',
   vertex 5 correspond the ticker '600010.SS',
   vertex 6 correspond the ticker '601098.SS',
   vertex 7 correspond the ticker '600005.SS',
   vertex 8 correspond the ticker '601225.SS',
   vertex 9 correspond the ticker '600372.SS',
   vertex 10 correspond the ticker '600519.SS',
   vertex 11 correspond the ticker '601336.SS',
   vertex 12 correspond the ticker '601186.SS'.
 
\end{itemize}

In the following table the number of rejected pairs of stocks from Chinese stock market for different $\alpha$(first row) and for different years (first column) are presented. For example the value 6 in the second column and second row of the table mean that in 2006 year only for six pairs of stocks from Chinese stock market individual hypotheses (\ref{individual_hypotheses}) were rejected. From the table one can see that for 2010 year and $\alpha=0.05$ hypothesis of elliptical model is not rejected.  

\begin{center}
\bf{Table 1.}
\begin{table}[h]
{
\begin{tabular}{|@{}|c|c|c|c|c|@{}|}\hline
year /$\alpha$ & 0.5 & 0.25 & 0.1 & 0.05 \\ \hline
2006 & 6 & 6 & 6 & 3 \\ \hline
2010 & 7 & 7 & 7 & 0 \\  \hline
2011 & 5 & 5 & 5 & 5 \\  \hline
\end{tabular} 
} 
\end{table}
\end{center}

\subsection{Testing of elliptical model for stock markets of different countries. Period of several years}

In the subsection the following question is considered: is it possible to accept the hypothesis of  elliptical model of stock returns distribution for the period from 2003 to 2014 despite on the fact that for some years from the period the elliptical model was rejected by the Holm procedure?

To answer this question, we propose the following procedure: the elliptical model is rejected if the number of years for which the elliptical model was rejected by the Holm procedure for one year period is greater than the given threshold.


The justification of the procedure can be given as follows. Let us formulate the hypothesis $H_i:\mbox{ 'in the year }i\mbox{ the distribution of stock returns is elliptical' }$. Then the number $X$ of rejected hypotheses  $H_i,i=1,\ldots,s$ for the period of length s is binomial random variable $b(s,\alpha)$ if all hypotheses $H_i,i=1,\ldots,s$ are true and Holm procedure for each year has FWER=$\alpha$. Here independence of decisions obtained by the Holm procedure for one year period follows from independence of observations for different years.  

Let $H=\bigcap_{i=1}^nH_i$ be the hypothesis that elliptical model for stock returns distribution is correct for all considered years. 
Then
$$P_{H}(X\geq c)=\sum_{i=c}^nC_n^i(\alpha)^i(1-\alpha)^{n-i}$$
Therefore for given $\beta$ one can define threshold $c_{\beta}$ from the equation
\begin{equation}\label{threshold_for_binomial}
P_{H}(X\geq c_{\beta})=\sum_{i=c_{\beta}}^nC_n^i(\alpha)^i(1-\alpha)^{n-i}\leq\beta
\end{equation}
and reject the hypothesis $H$ iff $X\geq c_{\beta}$.
This means that the test of the hypothesis $H$ has the form:
\begin{equation}\label{overall_hypothesis_test}
\varphi_H(x)=\left\{\
								\begin{array}{ll}
								  0,&x<c_{\beta}\\
								  1,&x\geq c_{\beta}
								\end{array}
							\right.	
\end{equation}
where $x$ is the observed number of the rejected hypotheses $H_i$, $c_{\beta}$ is defined from
(\ref{threshold_for_binomial}) and $\beta$ is the significance level of the test $\varphi_H(x)$.
If the Holm procedure was applied with FWER=$\alpha$ then the p-value of the test (\ref{overall_hypothesis_test}) can be calculated from:
\begin{equation}\label{p_value_overall_hypothesis_test}
p_{H_{\alpha}}(x)=\sum_{i=x}^nC_n^i(\alpha)^i(1-\alpha)^{n-i}
\end{equation}

Below we apply the procedure for stock markets of China, USA, Great Britain and Germany for the period from 2003 to 2014.
Presented experimental results contain the table with the pairs (i; j) of stocks for which the hypotheses $h_{ij}:{\cal Q}_{i,j}=\tau_{i,j}$ were rejected by the Holm procedure.
In all presented results the values $n=83,m=166$ were selected. Two different level of FWER for the Holm procedure was considered: $\alpha=0.05, \alpha=0.5$. 

\subsubsection{Chinese stock market}
For Chinese stock market (see table 2) the number of rejected hypotheses $h_{ij}:{\cal Q}_{i,j}=\tau_{i,j}$ is 11 for $\alpha=0.5$ and 10 for $\alpha=0.05$. From the other side 
the number of rejected hypotheses $H_i$ is equal to 5 for $\alpha=0.5$ and $\alpha=0.05$ (for years 2005, 2006, 2010, 2011, 2013). 


Then p-values (\ref{p_value_overall_hypothesis_test}) of the test $\varphi_H(x)$(\ref{overall_hypothesis_test}) of the hypothesis $H$ are 
$$
\begin{array}{l}
p_{H_{0.5}}(5)=P(N\geq 5|\alpha=0.5)=\sum_{i=5}^{12}C_{12}^i(0.5)^{12}=0.6128\\
p_{H_{0.05}}(5)=P(N\geq 5|\alpha=0.05)=\sum_{i=5}^{12}C_{12}^i(0.05)^{i}(0.95)^{12-i}=1.110779*10^{-5}
\end{array}
$$
One can see that p-value $p_{H_{0.05}}(5)$ is too small. Then hypothesis $H$ is rejected at any level $\beta>1.110779*10^{-5}$ if we test individual hypotheses $H_i:i=1,\ldots,12$ by Holm procedures with FWER $0.05$. In contrast hypothesis $H$ is accepted at any level $\beta\leq 0.6128$ if we test individual hypotheses $H_i:i=1,\ldots,12$ by Holm procedures with FWER $0.5$.    

\begin{center}
\bf{Table 2.}
\begin{table}[h]
{
\begin{tabular}{|@{}|c|c|c|@{}|}\hline
year  & $\alpha=0.5$ & $\alpha=0.05$ \\ \hline
2003 & 0 & 0 \\ \hline
2004 & 0 & 0  \\ \hline
2005 & (88;23) & (88;23) \\ \hline 
2006 & (4;2) & (4;2)  \\ \hline
2007 & 0 & 0 \\ \hline
2008 & 0 & 0  \\ \hline
2009 & 0 & 0  \\  \hline
2010 & (37;17),(85;66),(87;9),(92;56) & (37;17),(85;66),(87;9) \\ \hline
2011 & (22;5),(16;10),(19;16) & (22;5),(16;10),(19;16)  \\ \hline
2012 & 0 & 0 \\ \hline
2013 & (14;7),(29;2) & (14;7),(29;2)  \\ \hline
2014 & 0 & 0  \\  \hline
\end{tabular} 
} 
\end{table}
\end{center}

\subsubsection{USA stock market}
For USA stock market (see table 3) the number of rejected hypotheses $h_{ij}:{\cal Q}_{i,j}=\tau_{i,j}$ is 2 for $\alpha=0.5$ and 1 for $\alpha=0.05$. The number of rejected hypotheses $H_i$ is equal to 2 for $\alpha=0.5$ (for years 2004, 2006) and 1 for $\alpha=0.05$ (for year 2004). 
Then p-values (\ref{p_value_overall_hypothesis_test}) of the test $\varphi_H(x)$(\ref{overall_hypothesis_test}) of the hypothesis $H$ are 
$$
\begin{array}{l}
p_{H_{0.5}}(2)=P(N\geq 2|\alpha=0.5)=\sum_{i=2}^{12}C_{12}^i(0.5)^{12}= 0.9807\\
p_{H_{0.05}}(1)=P(N\geq 1|\alpha=0.05)=\sum_{i=1}^{12}C_{12}^i(0.05)^{i}(0.95)^{12-i}=0.1184
\end{array}
$$
Then hypothesis $H$ is rejected at any level $\beta>0.1184$ if we test individual hypotheses $H_i:i=1,\ldots,12$ by Holm procedures with FWER $0.05$ and hypothesis $H$ is accepted at any level 
$\beta\leq 0.9807$ if we test individual hypotheses $H_i:i=1,\ldots,12$ by Holm procedures with FWER $0.5$.    

\begin{center}
\bf{Table 3.}
\begin{table}[h]
{
\begin{tabular}{|@{}|c|c|c|@{}|}\hline
year  & $\alpha=0.5$ & $\alpha=0.05$ \\ \hline
2003 & 0 & 0 \\ \hline
2004 & (91;9) & (91;9)  \\ \hline
2005 & 0 & 0 \\ \hline
2006 & (59;22) & 0  \\ \hline 
2007 & 0 & 0 \\ \hline
2008 & 0 & 0  \\ \hline
2009 & 0 & 0  \\  \hline
2010 & 0 & 0 \\ \hline
2011 & 0 & 0  \\ \hline
2012 & 0 & 0 \\ \hline
2013 & 0 & 0  \\ \hline
2014 & 0 & 0  \\   \hline
\end{tabular} 
} 
\end{table}
\end{center}

\subsubsection{Great Britain stock market}
For Great Britain stock market (see table 4) the number of rejected hypotheses $h_{ij}:{\cal Q}_{i,j}=\tau_{i,j}$ is 1 for $\alpha=0.5$ and 0 for $\alpha=0.05$. 
The number of rejected hypotheses $H_i$ is equal to 1 for $\alpha=0.5$ (for years 2007) and 0 for $\alpha=0.05$. 
Then p-values (\ref{p_value_overall_hypothesis_test}) of the test $\varphi_H(x)$(\ref{overall_hypothesis_test}) of the hypothesis $H$ are 
$$
\begin{array}{l}
p_{H_{0.5}}(1)=P(N\geq 1|\alpha=0.5)=\sum_{i=1}^{12}C_{12}^i(0.5)^{12}= 0.9968\\
p_{H_{0.05}}(0)=P(N\geq 0|\alpha=0.05)=1
\end{array}
$$
One can see that both p-values are not too small. Then hypothesis $H$ is accepted at any level $\beta<0.9968$ if we test individual hypotheses $H_i:i=1,\ldots,12$ by Holm procedures with levels $0.05$ and hypothesis $H$ is accepted at any level $\beta$ if we test individual hypotheses $H_i:i=1,\ldots,12$ by Holm procedures with levels $0.5$.    
\begin{center}
\bf{Table 4.}
\begin{table}[h]
{
\begin{tabular}{|@{}|c|c|c|@{}|}\hline
year  & $\alpha=0.5$ & $\alpha=0.05$ \\ \hline
2003 & 0 & 0 \\ \hline
2004 & 0 & 0  \\ \hline
2005 & 0 & 0 \\ \hline
2006 & 0 & 0  \\ \hline
2007 & (57;29) & 0 \\ \hline
2008 & 0 & 0  \\ \hline
2009 & 0 & 0  \\  \hline
2010 & 0 & 0 \\ \hline
2011 & 0 & 0  \\ \hline
2012 & 0 & 0 \\ \hline
2013 & 0 & 0  \\ \hline
2014 & 0 & 0  \\   \hline
\end{tabular} 
} 
\end{table}
\end{center}

\subsubsection{Germany stock market}
 For Germany stock market (see table 5) the number of rejected hypotheses $h_{ij}:{\cal Q}_{i,j}=\tau_{i,j}$ is 2 for $\alpha=0.5$ and 0 for $\alpha=0.05$. 
The number of rejected hypotheses $H_i$ is equal to 2 for $\alpha=0.5$ (for years 2007) and 0 for $\alpha=0.05$. 
Then p-values (\ref{p_value_overall_hypothesis_test}) of the test $\varphi_H(x)$(\ref{overall_hypothesis_test}) of the hypothesis $H$ are 
$$
\begin{array}{l}
p_{H_{0.5}}(1)=P(N\geq 2|\alpha=0.5)=\sum_{i=2}^{12}C_{12}^i(0.5)^{12}=0.9807\\
p_{H_{0.05}}(0)=P(N\geq 0|\alpha=0.05)=1
\end{array}
$$
One can see that both p-values are not too small. Then hypothesis $H$ is accepted at any level $\beta<0.9807$ if we test individual hypotheses $H_i:i=1,\ldots,12$ by Holm procedures with levels $0.05$ and hypothesis $H$ is accepted at any level $\beta$ if we test individual hypotheses $H_i:i=1,\ldots,12$ by Holm procedures with levels $0.5$.    

\begin{center}
\bf{Table 5.}
\begin{table}[h]
{
\begin{tabular}{|@{}|c|c|c|@{}|}\hline
year  & $\alpha=0.5$ & $\alpha=0.05$ \\ \hline
2003 & 0 & 0 \\ \hline
2004 & 0 & 0  \\ \hline
2005 & 0 & 0 \\ \hline
2006 & 0 & 0  \\ \hline
2007 & (11;2) & 0 \\ \hline
2008 & 0 & 0  \\ \hline
2009 & (12;11) & 0  \\ \hline 
2010 & 0 & 0 \\ \hline
2011 & 0 & 0  \\ \hline
2012 & 0 & 0 \\ \hline
2013 & 0 & 0  \\ \hline
2014 & 0 & 0  \\   \hline
\end{tabular} 
} 
\end{table}
\end{center}

\section{Concluding remarks}\label{Concluding_remarks}

New property of elliptically contoured distributions namely the property of equality of $\tau$ Kendall correlation coefficient and measure ${\cal Q}$ of \cite{Kruskal1958} for any pair of random variables is proved in the paper. Distribution free individual tests for testing the property are constructed. Using well known Holm procedure these individual tests are combined to multiple hypotheses testing procedure.  

Application of the procedure to real market data from one year period shows that only small number of pairs of stocks from Chinese stock market destroy the tested property. 
Removing small number of stocks from consideration lead to nonrejection of elliptical model to the remaining stocks of stock market.
 
This result is consistent with the results obtained in \cite{Koldanov_2016} where symmetry hypothesis was tested. 
However in \cite{Koldanov_2016} the number of pairs of stocks which lead to rejection of symmetry hypotheses is huge but rejection graph has only small number of hubs (verteces with high degree).   
It is interesting to note than in contrast to \cite{Koldanov_2016} in the present case there are no hubs of high degree. For example in the figure \ref{Chinese_market_2006} there are only two hubs (verteces 3 and 8) of degree two. 

New procedure of testing the elliptical model for the period of several years is proposed. This procedure is applied for stock markets of China, USA, Great Britain and Germany for the period from 2003 to 2014. The obtained results shows that hypothesis of elliptical model of stock returns distribution is accepted for USA, Great Britain and Germany but is rejected for China.  

\section*{Acknowledgments}
This work is partly supported by Laboratory of Algorithms and technologies for network analysis of HSE and RFFI grant N 18-07-00524


\end{document}